\documentclass[12pt]{article}
\usepackage{amsmath,graphicx}
\usepackage{psfrag}
\usepackage{natbib}
\usepackage{url}

\makeatletter
\def\tagform@#1{\maketag@@@{[\ignorespaces#1\unskip\@@italiccorr]}}
\makeatother

\def \IR{\hbox{{\rm I}\kern-.2em\hbox{{\rm R}}}}
\newcounter{RSrefs}
\newlength{\RSdimenN}
\newlength{\RSdimenY}
\newlength{\RSdimenI}
\newcommand{\RSreference}[3]{(\stepcounter{RSrefs}\noindent
\hangindent=\the\RSdimenI\hangafter=1\hbox to
\the\RSdimenN{\arabic{RSrefs})\hss}\hbox to \the\RSdimenY{#1\hss}#2\par}

\begin{document}

\title{{\sc Julian Ernst Besag}\\ {\large 26 March 1945 -- 6 August 2010}\\ {\large Elected FRS 2004}
}
\author{ 
Peter J. Diggle\thanks {Lancaster Medical School,
Lancaster University,
Lancaster LA1 4YF, UK.
\newline \hspace*{5mm} Email: {\tt p.diggle@lancaster.ac.uk}}\\
University of Lancaster.\\
\and
Peter J. Green FRS\thanks {School of Mathematics, University of
Bristol, Bristol BS8 1TW, UK.
\newline \hspace*{5mm} Email: {\tt P.J.Green@bristol.ac.uk}.}\\
University of Bristol\\and University of Technology Sydney.\\
\and 
Bernard W. Silverman FRS\thanks {Department of Statistics,
University of Oxford,
24-29 St Giles',
Oxford OX1 3LB, UK.
\newline \hspace*{5mm} Email: {\tt bernard.silverman@stats.ox.ac.uk}}\\
University of Oxford.
}
\date{\today}
\maketitle

\begin{figure}
\begin{center}
\includegraphics[width=0.9\textwidth]{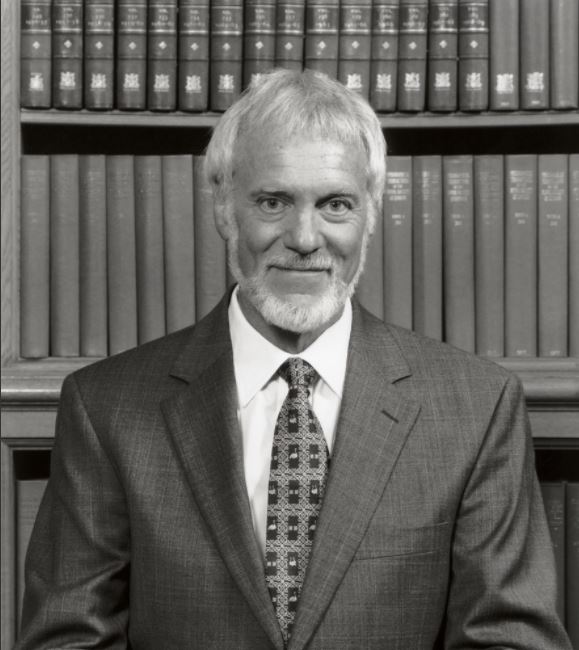}
\end{center}
\end{figure}

Julian Besag was an outstanding statistical scientist, distinguished for his pioneering work on the statistical theory and analysis of spatial processes, especially conditional lattice systems. His work has been seminal in statistical developments over the last several decades ranging from image analysis to Markov chain Monte Carlo methods. He clarified the role of auto-logistic and auto-normal models as instances of Markov random fields and paved the way for their use in diverse applications. Later work included investigations into the efficacy of nearest neighbour models to accommodate spatial dependence in the analysis of data from agricultural field trials, image restoration from noisy data, and texture generation using lattice models.
	
\section*{Family background, childhood and career}

\begin{figure}
\begin{center}
\includegraphics[height=0.35\textheight]{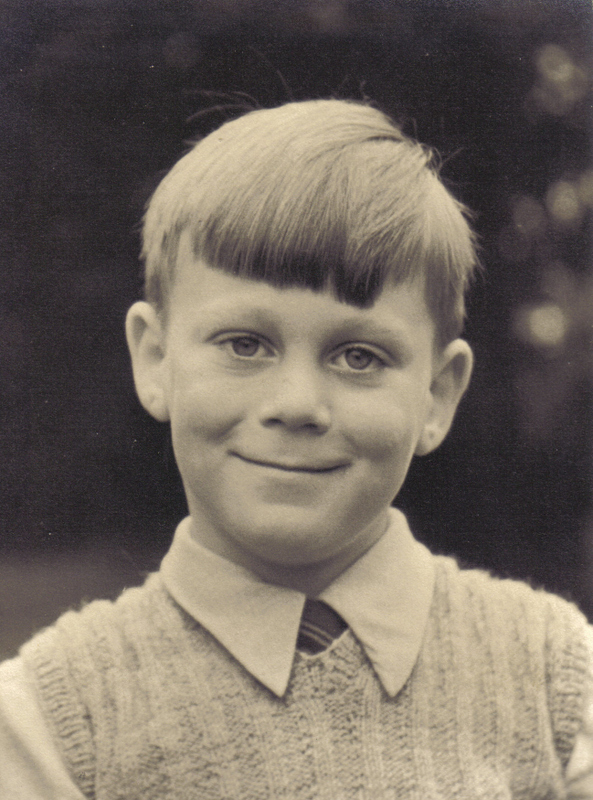}
\includegraphics[height=0.35\textheight]{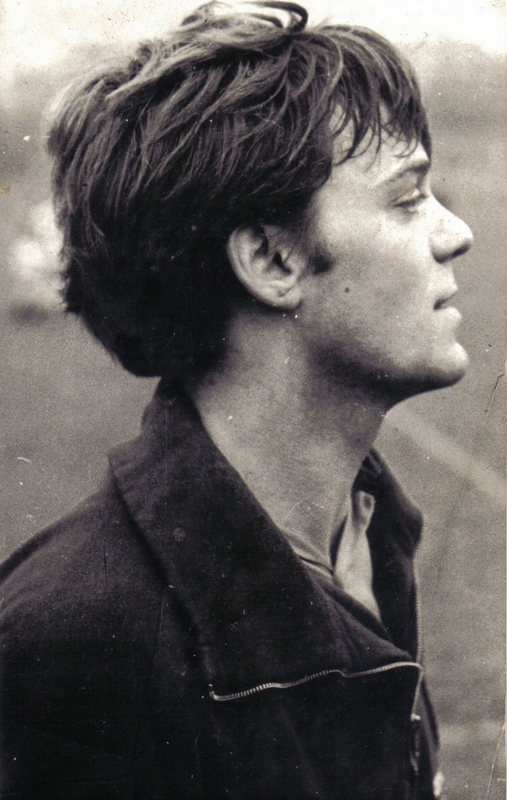}
\end{center}
\caption{Julian as a child, and as a hockey-playing student. \label{fig:child}}
\end{figure}

Julian Besag's father Emil Besag (1913--87) and paternal grandfather Ernst Besag (1878--1951) were both engineers born in Germany.   The Besag family was of Jewish origin, the name originating as a German version of the Hebrew name Pesach (Passover).   After his graduation from Munich in 1936, Emil moved to Birmingham to work for Crabtree Engineering; he was interned as an enemy alien at the start of the Second World War.   After his release he went to teach at Loughborough University,  where he remained for the rest of his career, becoming a professor specialising in techniques of measurement.   Ernst, who was the holder of a number of electrical equipment patents, worked until 1939 for a firm in Hornberg in the Black Forest which was linked by a licence agreement with Crabtree.   Shortly before the outbreak of war, Ernst also moved to work for Crabtree;  he was also interned but then returned to Crabtree with which he was linked to the end of his life.    Ernst's wife and four daughters remained in Germany and were deported from Baden to the Gurs concentration camp  in southern France in 1940.   In 1942 they were claimed as members by the Protestant church and were able to move to Switzerland for the remainder of the war.    
Julian's mother Irene (n\'{e}e Fuidge)  was a ballet dancer from Ireland.  She was already ill with cancer when Julian was born in Loughborough in March 1945, and she died in 1949.  He then moved to live with his German-speaking grandparents in Streetly, near Birmingham, moving back to his father's home when his father remarried in 1954 and subsequently becoming a pupil at Loughborough Grammar School; there are photographs of a youthful Julian in Figure \ref{fig:child}.   The Besags were enthusiastic mountaineers and Julian spent many family holidays climbing in the Alps. It is believed that in his teens he held the record for the longest survived fall from rocks in the UK.


Julian's early university career was unusual\footnote{This paragraph draws on the profile in the \emph{IMS Bulletin}, volume 33, issue 5 (2004)}. Following an unproductive period at Cambridge, nominally reading engineering, he obtained a BSc in mathematical statistics from Birmingham University in 1968. He considered himself ``immensely fortunate'' to have been there during a golden era in the Department of Mathematical Statistics, taking classes from Henry Daniels (chair, later FRS), Vic Barnett, Frank Downton, Ann Mitchell, John Nelder (later FRS), David Wishart and Steven Vajda. Julian then joined the Department of Biomathematics at Oxford University as full-time research assistant to Maurice Bartlett FRS, working on problems in the then fledgling subject of spatial statistics and what came to be  called Markov random fields. At the time, the Science Research Council would not allow its employees to register for a doctorate and so Julian's BSc remained his highest formal academic qualification throughout his career, a fact which he regarded as something of a badge of honour. 

\begin{figure}
\begin{center}
\includegraphics[width=0.6\textwidth]{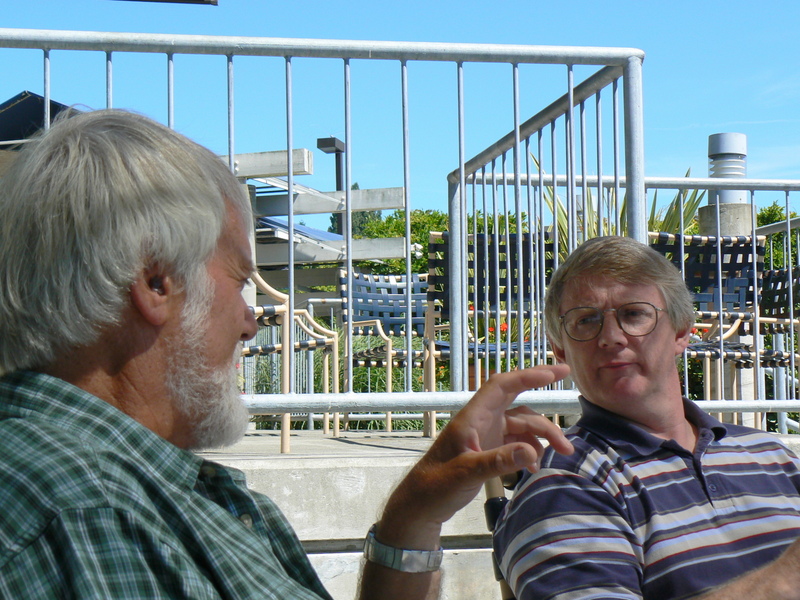}
\end{center}
\caption{Julian talking statistics with Chris Jennison. \label{fig:cj}}
\end{figure}

\begin{figure}
\begin{center}
\includegraphics[width=0.95\textwidth]{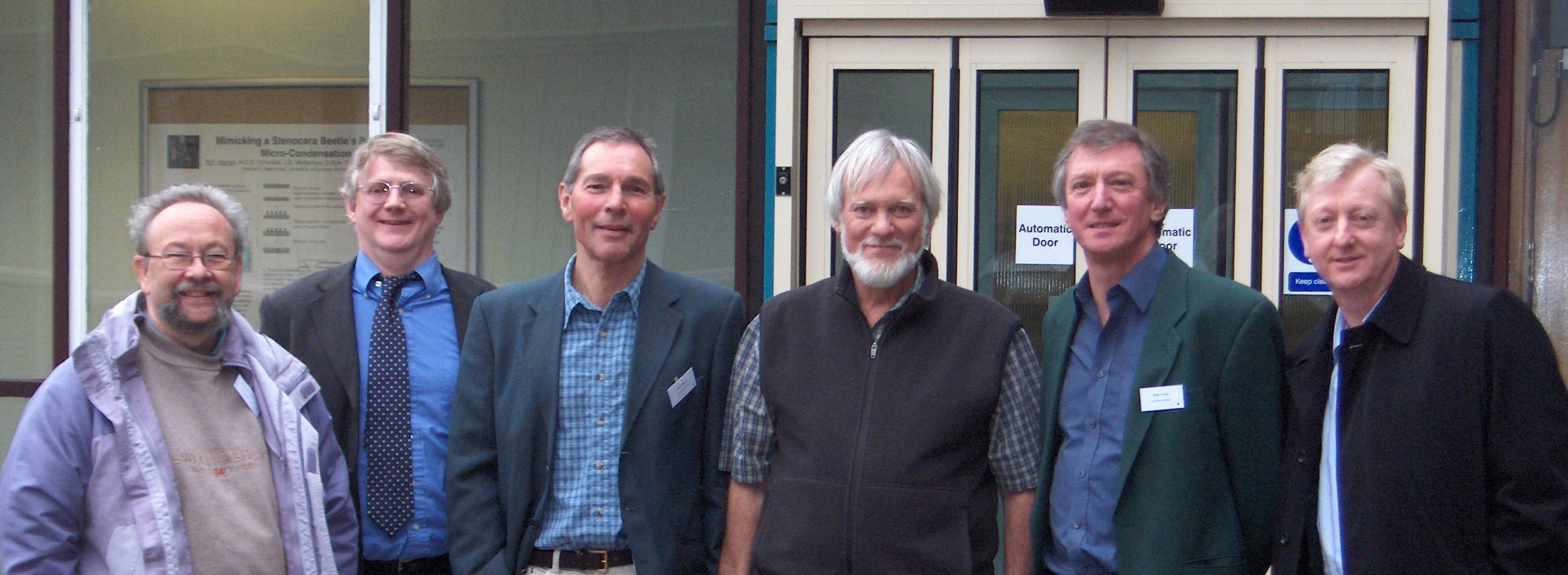}
\end{center}
\caption{Durham statisticians from the 70s on, reunited at a retirement event for Allan Seheult in 2007 -- from left, Alan Hawkes, Chris Jennison, Allan Seheult, Julian Besag, Peter Green, Bruce Porteous. \label{fig:durham}}
\end{figure}

\begin{figure}
\begin{center}
\includegraphics[width=0.8\textwidth]{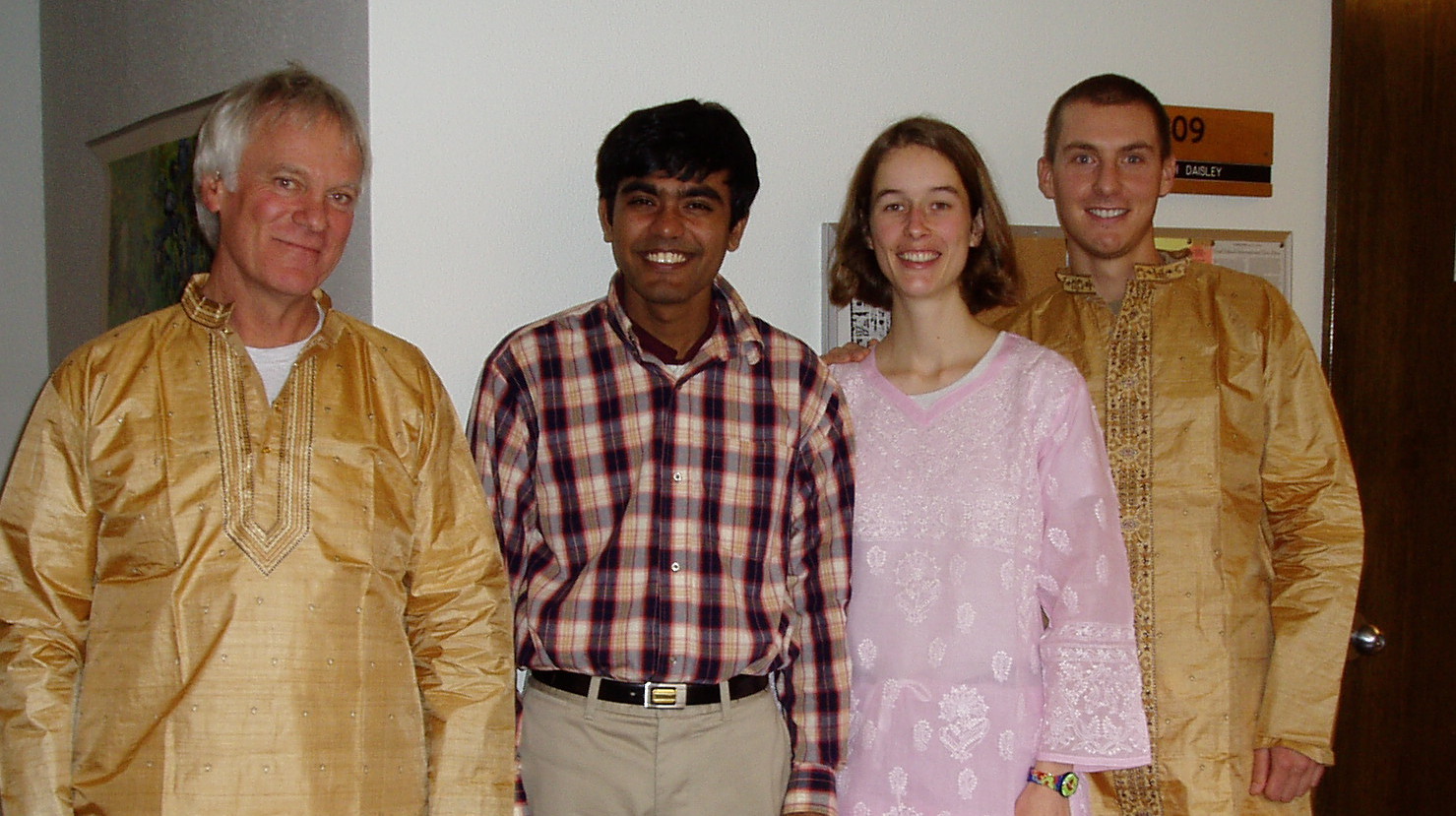}
\end{center}
\caption{Julian, Debashis Mondal, Marloes Maathuis and Raphael
Gottardo, at Debashis' defence of his PhD dissertation prospectus in Seattle. \label{fig:seattle}}
\end{figure}

He held a Lectureship at the University of Liverpool between 
1970 and 1975. 
After a very influential
sabbatical year spent visiting John Tukey (later ForMemRS) at Princeton, he moved to the University of Durham (see Figure \ref{fig:durham}) as
Reader until 1985 and as Professor from 1985 to 1989. Initially, Allan Seheult was the only other statistician there but Peter Green (later FRS) and Chris Jennison  (Figure \ref{fig:cj}) were subsequent recruits and the group flourished in spite of its small size. 
He then spent the 
academic year 1989/90 at the University
of Washington, Seattle, and 1990/91 at the University of Newcastle upon Tyne, before returning
to Washington (see Figure \ref{fig:seattle}) as Full Professor, a position he held until 2007. There he particularly enjoyed the wide range of
research interests and the recruitment of excellent junior faculty
and graduate students. He held a
Visiting Professorship at the University of Bath between 2007 and 2009, and was a visitor at
Bristol during the final year of his life. 

He married 
Valerie (n{\'e}e Brown) in 1966. Their son David was born in Chester hospital in 1972. Julian and Valerie became estranged in the 1980s, and Julian and David saw little of each other after that.
Julian remained very proud of them, especially of Valerie's books (on bullying) and of David 
as local councillor in Newcastle, and all three were reunited in the closing weeks of his life.

\section*{Scientific work}
Julian Besag's contributions to the discipline of statistics are profound, and continue to be of far-reaching consequence.
His research work had authority and great originality. He seldom wrote the last word on a subject, but was there at the start of many of the key developments in modern stochastic modelling and analysis. His record of publication is rather concise, but the work therein is very deliberate, and densely and painstakingly written. He had very high standards over what he put his name to.

\subsection*{Spatial statistics}
During the 1970's, work on the statistical analysis of spatially referenced data 
underwent a major expansion.   The major sources of scientific motivation included
forestry \citep[][but originally and remarkably published in 1960]{matern}, 
mining \citep{matheron1955application,matheron1963principles} and plant ecology \citep{bartlett1964spectral}. The applications in Julian's early work, including papers (1), (2) and (3),
were drawn from plant ecology and agriculture. The next six sections will explore Julian's work in spatial statistics and its applications in detail.

\subsection*{Modelling, conditional formulations}

Julian was a major contributor to modelling in the specific area that is
now usually labelled as discrete spatial variation. The ``discrete'' here refers to the space,
rather than to what is measured. Hence, the essence of the problem is to devise sensible 
models for a set of
random variables $Y_i:i=1,...,n$ linked to locations $x_i:i=1,...,n$ in two-dimensional 
space, and  associated methods of inference
for data consisting of a single realisation of $Y=(Y_1,...,Y_n)$. 

An obvious starting point is to assume that $Y$ follows a multivariate Normal distribution with a covariance matrix whose internal structure reflects the spatial context. One way to do this is
by a spatial analogue of the class of autoregressive models widely used in time series analysis,
hence
\begin{equation}
Y = A Y + Z,
\label{eq:SAR}
\end{equation}
where $Z \sim {\rm MVN}(0,\sigma^2 I)$. Equation [\ref{eq:SAR}] is called the
{\it simultaneous autoregressive} (SAR) model. A simple example is the first-order SAR on a regular square lattice of locations $x_i$, 
in which the element $a_{ij}$
of the square matrix $A$ is non-zero only when $x_i$ and $x_j$ correspond to
adjacent horizontal or vertical pairs of lattice-points.

Julian's early papers, culminating in the seminal (3), 
 read at an Ordinary Meeting of the Royal Statistical Society,
 argued persuasively that a more natural formulation
would be to define the joint distribution of $Y$ indirectly, via
a specification of the $n$ univariate conditional distributions of each $Y_i$ given the values of all other $Y_j$. From this perspective, the natural spatial analogue
of a time series autoregression is the {\it conditional autoregressive} (CAR) model,
\begin{equation}
Y_i|\{Y_j: j \neq i\} \sim {\rm N} (\sum_j b_{ij} Y_j,\tau^2): i=1,...,n.
\label{eq:CAR}
\end{equation}
The two classes of model  [\ref{eq:SAR}] and [\ref{eq:CAR}] are mathematically
equivalent in the
sense that any non-degenerate multivariate Normal distribution can be expressed in either form. From a
statistical modelling perspective, they differ in that what appears natural under one formulation
seems less natural in the other. Consider, for example, the following SAR on a one-dimensional
sequence of locations $x_i$,
$$
Y_i = a_1 Y_{i-1}+a_2 Y_{i+1} + Z_i.
$$
It seems reasonable to call this a {\it first-order} SAR, but it corresponds to the conditional formulation
$$Y_I|\{Y_j: j \neq i\} \sim {\rm N}\left(\sum_{j \neq i} b_{ij} Y_j,\tau^2\right),$$
where $b_{i-1}$ and $b_{i+1}$ are non-zero as expected,  but $b_{i-2}$ and $b_{i+2}$
are also non-zero, i.e. this is a {\it second-order} CAR.

A more fundamental advantage of the CAR over the SAR formulation is that it extends naturally to 
non-Gaussian models. Specifically, the linear formulation of [\ref{eq:CAR}] can be replaced by the
analogous generalized linear formulation to define a spatial counterpart of the class of
generalized linear models for independent $Y_i$ introduced by
\citet{nelder}. Indeed, Julian's first explicit proposal (1) was a logistic
model for spatially referenced binary $Y_i$. 
The downside to this apparent
gain in flexibility was the need to ensure the self-consistency of the $n$ univariate
conditional distributions of each $Y_i$ given all other $Y_j$. This obstacle was removed,
or rather its precise nature  made clear, by the celebrated Hammersley--Clifford theorem -- 
celebrated both for its importance and for its never having been published by its originators Peter Clifford and John Hammersley (later FRS).
Julian's paper (3) included a simple proof of the theorem and demonstrated its use
both in constructing useful models and in
explaining why the CAR formulation was not a panacea; one of his examples
showed that an apparently natural construction 
for a model with Poisson conditionals was valid only if its spatial 
dependence parameters
were constrained to impose a negative dependence between counts $Y_i,Y_j$ at 
neighbouring locations $x_i,x_j$.

The influence of (3) spread far beyond its specific context. In seconding 
the vote of thanks, Alan Hawkes made the prescient remark that the paper 
contained an ``elegant general treatment of distributions on lattices -- or, 
indeed, {\it for any multivariate distribution at all}'' (our italics). Hawkes' observation
preceded by six years the first paper on the now-ubiquitous class of 
{\it graphical models} \citep{darroch1980markov}, which drew not on the 
spatial statistics literature but on the theory
of Markov fields \citep{kemeny1976denumerable} applied to models for 
multidimensional contingency tables. 
\begin{figure}
\begin{center}
\includegraphics[height=0.45\textheight]{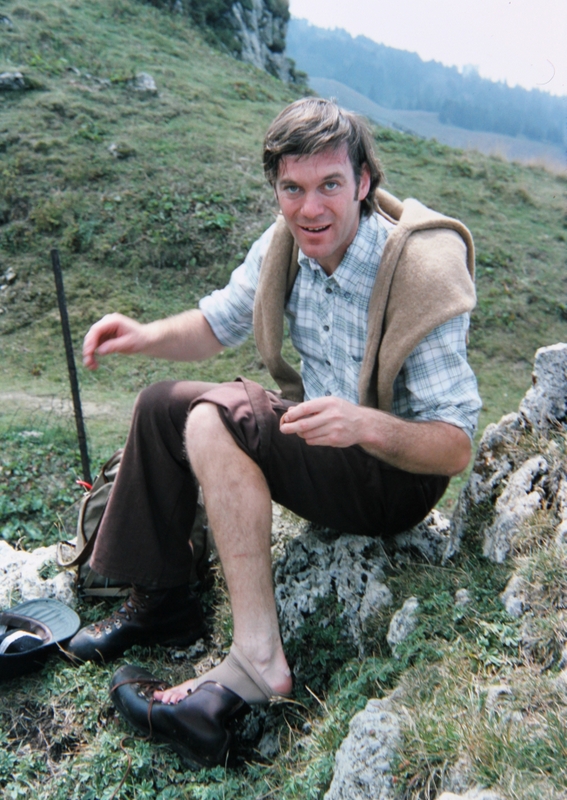}
\includegraphics[height=0.45\textheight]{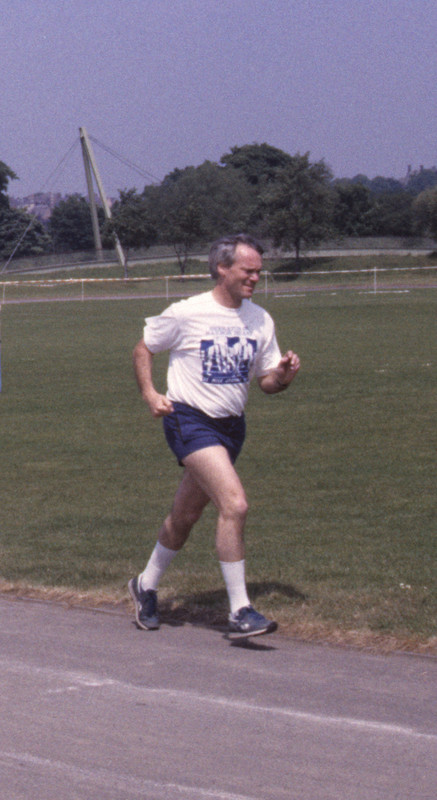}
\end{center}
\caption{Julian on a hike in the Vercours in 1976, and in the 1989 edition of the Durham annual \emph{Examiner's mile}. \label{fig:sport}}
\end{figure}

\subsection*{Approaches to inference}

Whilst Julian's preferred conditional formulation of spatial models
opened the door to a range of non-Gaussian formulations, the
intractability of their associated joint distributions 
made for
an awkward inferential problem, since standard likelihood-based
methods of estimation and testing were not readily available.

In (3), Julian's solution was a {\it coding} method.  Because the
conditional distribution of each $Y_i$ given all other $Y_j$ typically
depended only on a subset of the $Y_j$, it was possible to code
each of the locations  $x_i,i=1,...,n$ as black or white in such a way
 that the joint distribution
of the $Y_i$ at all of the black locations, conditional on their values 
at all of the white locations, was a simple product of explicit univariate
conditional distributions, that can be used legitimately
for likelihood-based
inference.
For example, when the locations form a square
lattice and the neighbours of $(i,j)$ are $(i,j-1), (i,j+1), (i-1,j)$ and
$(i+1,j)$ the coding method results in a chess-board pattern and two,
equally valid, likelihoods result by reversing the roles of the
black and white 
locations. A simple way to pool the information is to take the product of
the two, which gives a special case of what Julian later proposed
more generally in (4) 
as a {\it pseudo-likelihood}  for any 
multivariate distribution of random variables $Y_1,...,Y_n$, namely
the product of the $n$ univariate conditional distributions of each 
$Y_i$ given all other $Y_j$. 

The intractability problem for conditional model specifications
was later circumvented by the advent of Markov chain Monte Carlo (MCMC) methods in statistics,
which apply very naturally to models of this kind.
Together with the development of powerful and easy-to-use
software such as the BUGS system \citep{gilks}, MCMC methods were
instrumental in the explosive growth of Bayesian methods in
applied statistics since the 1990s. Julian's approach to
inference undoubtedly moved towards the Bayesian paradigm over time,
but was always tempered by a strong streak of pragmatism. In opening
the discussion (23) 
of \citet{mccullagh}, he remarked with reference to (21) 
that ``David Higdon and I are primarily `spatialists' rather 
than card-carrying Bayesians.'' 

Throughout his career, Julian thought deeply about the role of
statistical modelling in  {\it scientific} inference.  In this
context, he considered
the term ``model'' itself to be grossly over-used. He thought
that it should be reserved for mathematical formulations with
a direct connection to an underlying mechanism, citing as
an example the Poisson process as a model in radiation physics
(23). He argued unsuccessfully for the term ``statistical
model'' to be replaced by ``statistical scheme'' and was
always acutely aware, in the spirit of George Box's well-known
aphorism, that ``all models are wrong but some are useful''
\citep{box}. He was not reluctant to criticise other points
of view, for example in (23), 
but he subjected his own work
to intense self-criticism. 

\subsection*{Algebra of interacting systems}

Much of Julian's research output on Markov random fields was focused rather directly on delivering new statistical methodology for spatial statistics, but he also published a series of papers exploring more fundamental mathematical aspects of these fields; this series is quite sparse, and spread through his career. 

A common strand in many of these papers was the behaviour of CAR processes approaching the non-stationary limit, a region of the parameter space of particular importance in many applications, although one posing some additional mathematical technicalities. Quite early in his career, paper (6) 
explored the auto-correlation structure of stationary auto-normal schemes on an infinite square lattice. Julian notes the very slow decline in auto-correlation with distance, close to the limit. In the isotropic case (north-south dependence structure the same as east-west), he hints at a connection to continuous-space isotropy, something that he only resolved successfully much later in his career. 

Paper (19) with Kooperberg 
provides further theory on conditional and intrinsic autoregressions; the main contribution is a partial synthesis  of standard geostatistical and Gaussian Markov random field
formulations. This theme is taken further in (24), 
where the focus of interest is the limiting continuous-space behaviour of these Gaussian fields as the spatial discretisation becomes finer. The object of study is a Gaussian intrinsic first order two-dimensional autoregression, where the limit is the de Wijs process, a generalised process; the nature of this limit is analogous to the way in which Brownian motion is obtained as the limit of a random walk in one dimension. 

Earlier, in another of his most mathematically-intense papers, he looked at some discrete-state analogues. The setting in (8) 
is that of auto-Poisson lattice schemes, where it is shown that (almost) any purely inhibitory pairwise-interaction point process can be obtained in the limit. Other pairwise-interaction processes are obtained as limits of sequences of auto-logistic lattice schemes. 

Some important further contributions to discrete-state Markov random fields include (20), 
which examines processes with higher-order interactions, and shows that more satisfactory posterior probabilities on image features can then be obtained, compared to the standard nearest-neighbour approach, and (22), 
which gives some new insights into exact computations with Markov random fields. 

\subsection*{Digital image analysis}

Although not anticipated at the time of the  appearance of Julian's 1974 paper (3), a very natural domain where Markov random fields on a rectangular lattice would prove an extremely influential source of models was digital image analysis. Developments in technologies such as satellite remote sensing and medical imaging, together with increase in computer power, meant that by the early 80s the idea of storing images -- whether formed from visible light or based on other kinds of radiation such as sonar or radar reflections, or emitted photons -- as rectangular arrays of digitised intensity values was becoming commonplace, and electrical engineers and computer scientists were becoming active in devising methodologies for processing such images. The objective at this early stage was very much on separating signal from noise.

These researchers often had a background in signal processing, and their knowledge of statistical models was focused especially on time series processes, so they were naturally inclined to see images as indexed by two-dimensional time, and to seek two dimensional analogues of models they were familiar with in much the same way, and with the same motivation, as statisticians and probabilists had done earlier. 

By the early 80s, paper (3) was becoming cited in the image analysis literature, and so this new science had come to his attention, but the work that really stimulated his curiosity and interest was that by Don and Stuart Geman. Their seminal paper \citep{geman1984stochastic} was available a few years before publication, and Don Geman visited Durham in about 1983; this work aroused intense interest among Julian and his colleagues. The key development that opened up the field subsequently known as Bayesian image analysis was the idea of using a spatial stochastic process such as a Markov random field not for the observed image (rectangular array of intensities) but as a model for an unobserved `true' image or `ground truth', of which the observed image was a noisy version. Adopting this view was liberating, as it allowed for separating the concerns of realistically modelling the observed intensity distributions from that of capturing the spatial dependence across the elements (pixels) of the image. Thus the Geman and Geman work was based on a discrete-valued Markov random field for the ground truth coupled with a noise process that might with equal facility and tractability be modelled as binary noise or using a continuous distribution such as the normal. 

Julian's first published contribution to the field was his provocatively-titled 1986 paper `On the statistical analysis of dirty pictures' (9), read to the Royal Statistical Society. 
For the most part, the dirty pictures studied are noise-corrupted discrete-valued fields, modelled as Markov random fields as in \citet{geman1984stochastic}, so naturally viewed as spatial hidden Markov models. Interestingly, the language of the paper is to regard the unobserved true image (`ground truth') as a realisation of a stochastic process, so this is a latent variable problem. By the time the rejoinder to the copious discussion is composed some months later, however, Julian is using explicitly Bayesian language. This employs a formulation in which the ground truth is a parameter in a Bayesian setting, with the random field model as its prior, and in which the goal is to use the resulting posterior for the inference that represents reconstruction of the image. 

Whatever the formal paradigm, the key contributions of the paper are the careful examination of the use of MRF models as image priors, in particular the distinction between their small-scale and large-scale properties, a parallel comparative study of local and global methods of reconstruction, and a discussion of the difficulties of estimating (hyper-) parameters. As a constrast to the extant approaches of posterior sampling (using for example the Gibbs sampler newly coined by \citeauthor{geman1984stochastic}) to obtain posterior means or marginal posterior modes of the true image, and the use of simulated annealing to approximate the \emph{maximum a posteriori} (MAP) image estimate, Julian proposed his `iterated conditional modes' algorithm, essentially a Gauss--Seidel approach to optimisation in this discrete variable context, with true pixel values sequentially updated to maximise the local characteristic (equivalently, potential or full conditional). Experiments demonstrated improvements in empirical reconstruction performance compared to MAP, as well as very substantial computational savings.

Julian's later contributions to statistical image analysis adopted Bayesian language more consistently, including (11) and (15).
The former investigated fully Bayesian image analysis, including hyperpriors, implemented by MCMC; the latter demonstrating applications outside digital image analysis literally, to spatial problems of similar formal structure including in archaeology and epidemiology. Although his focus was by then less on devising reconstruction methodology, later work including (19) and (20)
was very much inspired by the goal of constructing richer classes of prior for image models.

\begin{figure}
\begin{center}
\includegraphics[height=0.25\textheight]{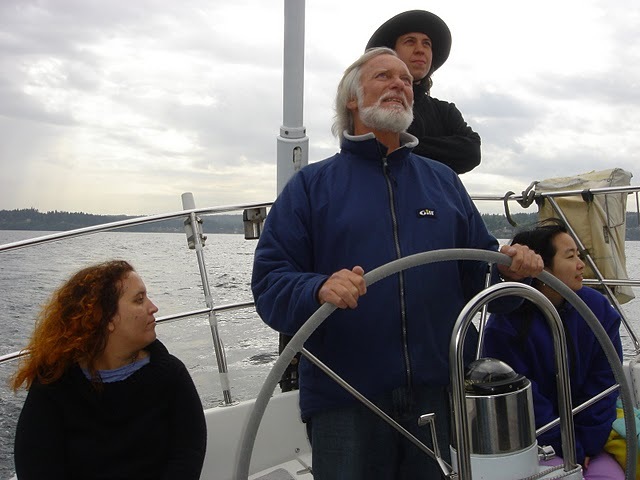}
\includegraphics[height=0.25\textheight]{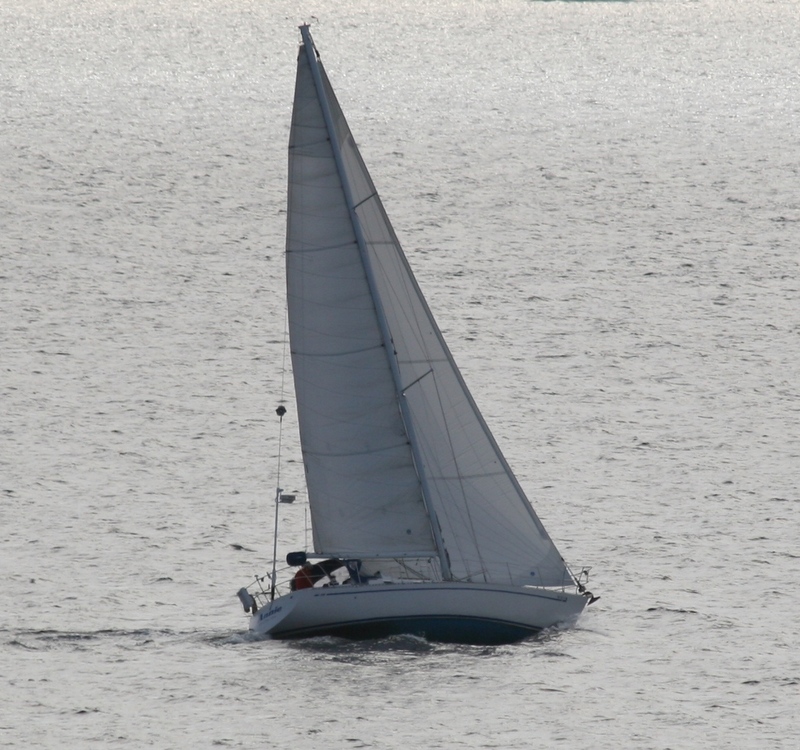}
\end{center}
\caption{Julian at the helm of his sailboat \emph{Annie}, seen under sail on right. \label{fig:annie}}
\end{figure}

\subsection*{Epidemiology}

Epidemiology provided perhaps the earliest documented example of what we would now call
a spatial analysis in the form of John Snow's famous map of cholera deaths in Soho, London, 
during the 1854 epidemic. This convincingly demonstrated the link to a contaminated
public water supply from the Broad Street pump; for a detailed discussion, see 
\citet{hill1955} or
\citet{cliff1992atlas}. Despite this head-start, epidemiology played a rather
minor role in the development of mainstream spatial statistical methodology in
the 1960s and 1970s. In a series of papers, 
\citet{knox1963detection,knox1964epidemiology} \citep[see also][]{knox1964detection} proposed a
statistical test for space-time interaction that he applied to data on the 
spatio-temporal distributions of cleft lip and palate, and of
childhood leukaemia cases. Bartlett appears not to have followed up this
particular area of application. Had he done so, Julian would presumably 
have become involved when he moved to Oxford in 1968, rather than in Newcastle upon Tyne some 
20 years later. 

The link between spatial statistical methods and epidemiological applications was 
re-established during the 1980s, in two somewhat different settings: disease clustering and disease mapping. 

 Julian's year in Newcastle upon Tyne (1990--91) coincided with a growing
 interest in the UK in using spatial statistical methods to detect unusual 
 concentrations, so-called ``clusters,''
 of cases of a rare disease. This was in part stimulated by the Black 
 inquiry \citep{black}
 into a reported excess of cases of childhood leukaemia in the vicinity of the
 Sellafield nuclear power plant in West Cumbria. For example, the Black report
 led to the release of national government funding to establish a
  ``Small Area Health Statistics Unit'' under the direction of Paul Elliott,
 initially in the London School of Hygiene and Tropical Medicine but now in 
 Imperial College. 

 Around that time, the Newcastle geographer Stan Openshaw
 had developed a method for cluster detection that involved simultaneous
 testing for evidence of clusters over a partition of a geographical region 
 into many small areas \citep{craft,openshaw}. Julian admired this work but felt that it lacked a rigorous
 statistical foundation, which he then provided with Newell in (14).

A key early paper on disease mapping was \citet{clayton1987empirical}. The essence of the problem
is to estimate spatial variation in disease risk from data consisting of the numbers 
of cases, $Y_i$ and numbers at risk, $n_i$, in each of $n$
sub-regions $A_i$ that form
a partition of the region of interest, $A$.
\citet{clayton1987empirical} proposed a hierarchical statistical model in which, conditional
on the risks, $R_i$, associated with each $A_i$, the $Y_i$ are mutually independent Poisson
random variables with conditional means $n_i R_i$. They considered both an exchangeable,
non-spatial
model in which the $R_i$ were mutually independent gamma or log-Normal variates, and a spatial
 CAR model, in which the full conditional distributions of
$B_i = \log R_i$ are Normal, with conditional means
$$
{\rm E}[B_i|B_j: j \neq i] = \mu_i + \rho \sum_{j \neq i} W_{ij} B_j
$$
and common conditional variance ${\rm Var}(B_i|B_j: j \neq i) = \sigma^2$. Here,
$W_{ij}=1$ when $A_i$ and $A_j$ share a boundary segment, $W_{ij}=0$ otherwise.

Julian recognised the practical importance of the Clayton and Kaldor paper but 
questioned their detailed
 formulation of the CAR model, which he referred to as 
involving ``a slight logical inconsistency'' (15).
The remedy was to replace the constant conditional variance 
$\sigma^2$ by
$${\rm Var}(R_i|R_j: j \neq i) = \sigma^2/n_i$$
where $n_i$ denotes the number of {\it neighbours} of $i$,
i.e, in the present context the number of $A_j$ that share a boundary segment with $A_i$;
the 1991 paper with York and Molli\'{e} (15) had a great impact. The eponymous BYM method
has become the standard approach to disease mapping. More than this, it was an early example
of the now-ubiquitous hierarchical formulation of spatially dependent models for discrete data.

\subsection*{Agricultural field trials}

One common setting in
which discrete spatial variation arises naturally is that of agricultural field trials,
where $Y_i$ represents the yield of the $i$th of $n$ plots and $x_i$ a convenient reference
point for the position of the $i$th plot within the overall layout.
R. A. Fisher (later Sir Ronald Fisher FRS) contemplated
a model-based approach to the analysis of field trials 
when working at the Rothamsted experimental station in the 1930's. In {\it The Design of
Experiments} \citep{fisher1937design} he commented on ``the widely verified fact that patches in close
proximity are commonly more alike, as judged by the yield of crops, than those which are farther
apart.''  
As is well-known, Fisher addressed this by advocating 
randomised block designs
in which each block is a set of contiguous plots, and the induced randomisation
distribution is the basis for inference. 

More than 40 years later,
the integration of spatial statistical methods into the analysis of 
agricultural field trials was stimulated by two papers read at Ordinary Meetings
of the Royal Statistical Society \citep{bartlett1978nearest,wilkinson1983nearest}. 
Bartlett's paper was primarily a re-assessment, in the light of later theoretical
developments in spatial statistics, of an empirical device proposed
by \citet{papadakis1937methode} whereby plot yields are adjusted 
by treating the yields on neighbouring plots as covariates. Bartlett viewed
the Papadakis adjustment as ``an ancillary exploratory device.''
In his contribution to the
discussion, Julian set out  the elements of an overtly spatial model-based approach, within
which the parameters of interest could be estimated by maximum likelihood whilst 
incorporating an explicit parametric
 model for the spatial correlation between neighbouring plot yields. Characteristically,
in proposing a model-based solution
Julian acknowledged its attendant risks: 
``presumably, one cannot appeal to randomization arguments when 
assumptions are materially violated.'' 

Julian followed up this suggestion by visiting Rob Kempton at the
Plant Breeding Institute in Cambridge, UK, in 1982. Their collaboration
led to a paper (10) 
that discussed, through a series
of carefully chosen case-studies, the need for the approach to spatial analysis of
field experiments to reflect the
different ways in which 
spatial dependence might arise: spatial variation in soil fertility; competition between
plants in neighbouring plots; leakage of treatments across plot boundaries.  The paper
also recognised the tension between model-based and design-based approaches to
inference in field experiments, commenting that ``It is salutary to note that a large
proportion of experiments throughout the world are conducted without sophisticated design''
and that as a consequence
``the statistician, whilst presumably giving advice on better designs for future experiments, is often faced with salvaging a current trial.'' 

This same tension was evident when, some years later, Julian and 
a University of
Washington PhD student, David Higdon, published 
a Royal Statistical Society read paper (21) 
that
set out a comprehensive framework for the analysis of field trials, which 
combined spatial modelling and Bayesian inference. Not all of the discussants of
the paper were convinced by the arguments put forward in the paper. In particular,
the explicitly model-based approach was felt by some to be at odds with the strong tradition in favour of the design-based approach.

An intriguing footnote is that spatial statistical methods for analysing field trials 
seem to have been adopted most enthusiastically in Australia, where their use has become
widespread: see, for example, \citet{williams1986neighbour}, \citet{gilmour1997accounting},
\citet{cullis1998spatial} and \citet{stefanova2009enhanced}. This may have at least
as much
to do with its promulgation by prominent Australian statisticians working within 
Australian government-sponsored rural industries research agencies as with anything
uniquely different about Australian fields. However, it is also the case that
field experiments conducted on a large physical scale with many plots
are more amenable to spatial modelling, and experiments of this kind
are typical in
several areas of Australian agriculture.

\subsection*{Markov chain Monte Carlo methods}

Julian was well ahead of his time in recognising the duality between the conditional specification of stochastic models and the construction of algorithms for such models using these conditional specifications. One notable instance of this was the `iterated conditional modes' algorithm for finding local modes of posterior distributions, proposed in the context of image analysis in his `dirty pictures' paper (9); 
see \emph{Digital image analysis}, above. More importantly, this led him to study Markov chain Monte Carlo methods. He adopted Ulf Grenander's maxim that `pattern analysis equals pattern synthesis' 
\citep[see][in work actually widely available before that]{grenanderlectures}.

So far as we know, he did not explicitly notice that simulating from full conditional distributions could be a completely general tool for sampling from posteriors for Bayesian models and other complex systems, and like almost everyone else in statistics, he missed the significance for general statistical methodology of the landmark papers by \citet{metropolis1953equation} and \citet{hastings1970monte}. Furthermore, he continued to focus attention on spatial models alone until the very late 80s. So it is 
\citet{gelfand1990sampling} 
who obtained and deserved the credit for initiating the burst of interest of MCMC methods in Bayesian statistics from about 1990. 
Julian didn't see this coming in for example (9) and (11), 
yet in 1975, he had written (4): 
\begin{quotation}
In section 3, we shall be discussing  some methods of statistical  analysis
for conditional  probability  schemes.  Since the sampling  properties  of the
techniques,  beyond consistency,  are largely unknown and likely to be
analytically  intractable,  it would be useful to carry out Monte Carlo
simulation  studies, where feasible. For discrete random variables, no
direct methods of simulation  have yet been found. In principle,  it is
possible to set up a discrete  time, spatial-temporal  Markov chain which
yields as its stationary  temporal  limit the required spatial distribution.
The simulation  procedure  is to consider the sites cyclically  and, at each
stage, to amend or leave unaltered  the particular  site value in question,
according  to a probability  distribution  whose elements  depend upon the
current  values at neighbouring  sites. For further  details, see \citet[Chapter 9]{hammersley5monte}; however, the technique  is unlikely  to
be particularly  helpful  in many other  than binary  situations and the Markov
chain itself  has no practical interpretation.
\end{quotation}
This apparently anticipated not only the Gibbs sampler, usually attributed to \citet{geman1984stochastic}, but also more general (single-variable-update) MCMC methods. As noted by \citet{robert2011short}, he was ``clearly understating the importance of his own work.''

Nevertheless, Julian deserves his reputation as one of the very early proponents of Markov chain Monte Carlo methods for fitting statistical models (15, 16 and 18). 
These delivered interesting and novel methodologies for challenging inferential problems in various application domains, such as spatial archaeology, spatial epidemiology, age--period--cohort models, agricultural variety trials, medical imaging. Apart from this, the last two papers were key, in the early years of MCMC's re-discovery, in helping our collective understanding of the potential of MCMC in statistical model fitting move beyond the Gibbs sampler to other sampling methods, including the Metropolis and Langevin algorithms.

Along with explicating and promoting MCMC methodology in its early years in statistics, and bringing a number of ideas from the statistical physics literature into statistics, Julian's MCMC work introduced a number of innovations to MCMC methodology itself.
He proposed a `Langevin--Hastings algorithm' (later known as the Metropolis-adjusted Langevin algorithm (MALA)) in the 
RSS discussion contribution (17). 
Paper (18), with Green, Higdon and Mengersen, 
introduces MCMC methods based on partial (not full) conditioning; this gives a general framework for integrating prediction with inference in statistical models, and extends multigrid Swendsen--Wang. Other innovations include randomised proposals -- which explain adaptive rejection Metropolis sampling -- and a method for simultaneous credible regions, for inference about functions and surfaces. 

\subsection*{Monte Carlo hypothesis testing}

The use of Monte Carlo methods to enable inference with
analytically intractable models is now thoroughly integrated into statistical
practice but was novel in the 1970s. The idea of a simple Monte Carlo test, which in essence is equivalent to sampling from, rather than complete enumeration of, the randomisation distribution of a test statistic under the null hypothesis of interest, was foreshadowed in Fisher's writing, made explicit by \citet{barnard} 
and used extensively in \citet{ripley}
to assess the goodness of  fit of spatial point process models. Paper (5) with Diggle 
is a collection of examples in a variety of spatial settings, written to promote the wider use of Monte Carlo testing as a tool in exploratory data analysis.

In more challenging settings, where direct simulation under the null hypothesis is not practical, it is tempting to consider MCMC instead. Remarkably, Julian was able to demonstrate that exactness of computed p-values can be retained even with MCMC. The setting is goodness-of-fit tests for those models where the likelihood for data $x$ depends only on the value of a sufficient statistic $T(x)$. In this situation, the distribution of the data given $T(x)$ is uniform, as observed by Fisher. This immediately generates a Monte Carlo goodness-of-fit test for the assumed model; for any suitable test statistic, the value $U(x_\text{obs})$ calculated from the data is compared to the distribution of $U$ values calculated from a sample from $x$ given that $T(x)=T(x_\text{obs})$. The role for MCMC here is to greatly extend the range of complex models where this is usable, since we can use a MCMC sampler targeting the conditional distribution instead of a direct sample, if it is used in the right way.

In (12), 
Julian showed two ways to guarantee immunity to transient MCMC bias and effects
of dependence, so that p-values are exact -- a `serial' and a `parallel' method. For a concrete example, a goodness of fit test for an Ising model conditional on the number of `black' sites can be constructed by running a Markov chain for this conditional model \emph{backwards} some number $N$ of steps and then from the resulting state running forwards $N$ steps, repeating this forward simulation from this same initial state multiple times. The value of any test statistic calculated on the data is exchangeable with those of the simulated replicates, which is enough to ensure exactness of the resulting p-values. This provided the first tractable approach for the Ising and Rasch models. Paper (13), also with Clifford, 
showed how to control computational costs in this kind of scheme, in both regular Monte Carlo testing and the MCMC variants, with a sequential approach that 
can truncate the simulation; repeat the generation of test statistic values until either $n$ values have been obtained, or
$h$ values exceed that of $u=U(x_\text{obs})$, and report a (conservative) p-value of $(g + 1)/n$
where $g$ is the number of values exceeding $u$ on termination -- with great saving on computing when the p-value is large.

At the time of his death, Julian left unfinished some work with Debashis Mondal, which Debashis later completed and published as (25), in a paper showcased by \emph{Biometrics} at the Joint Statistical Meetings in 2014. 
This extends the scope for Monte Carlo-based goodness of fit testing yet further, with a portfolio of ingenious tricks handling various classes of Markov chain. 
\begin{figure}
\begin{center}
\includegraphics[width=0.6\textwidth]{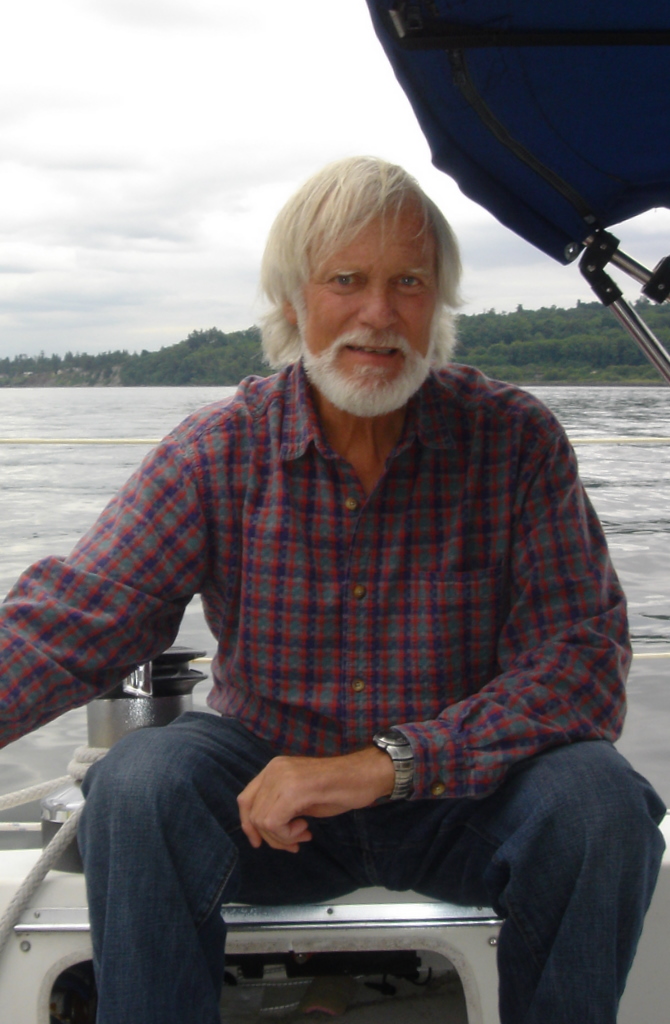}
\end{center}
\caption{Julian on the water, close to Seattle. \label{fig:helm}}
\end{figure}

\subsection*{Exploratory Data Analysis}
One of the major areas of development and thought in Statistics in the 1970s and early 80s was Exploratory Data Analysis (EDA).  The focus of this was J. W. Tukey at Princeton.  Tukey had started life as a topologist, but had been involved with computers since the 1940s, which is noteworthy since his instrument of choice for working on data was a multi-coloured ball pen!   Exploratory Data Analysis as expounded by Tukey was, in essence, a whole philosophy which reacted against more formal and mathematical approaches to statistics.  His 1977 book Exploratory Data Analysis \citep{Tukey:1977:EDA}, the seminal text, almost the bible, of the field must be the only book intended for statistics undergraduates which nowhere suggests the notion of taking the average of a collection of numbers.  Crucial to Tukey's thinking was to use measures such as the median and to look at outlying observations individually.

Julian was one of the older members of a community of British \lq\lq{}young statisticians\rq\rq{} that included the three authors of this memoir, which often met in the pub after Royal Statistical Society discussion paper meetings (and at other times too).  Julian often contributed to the discussions of the papers himself and 
encouraged others to do so. It is not surprising that one of his discussion contributions, to the paper by Ripley (1977), has been cited over 500 times. He set an 
example of genuine intellectual curiosity and the desire to get to the 
truth of the matter through straightforward discussion. His public 
approach during the meetings themselves spilled over into all 
the informal conversation afterwards. And of course the networks started 
off at Royal Statistical Society meetings grew into lifelong friendships and working relationships.

Julian was very strongly influenced by the year he spent at Princeton at that time, and he and the other young statisticians were very interested in EDA.  Julian was both extremely enthusiastic about EDA but also prepared to be critical of its excesses.  For example, Tukey (who was greatly given to coining names for things, and is credited with \lq\lq{}bit\rq\rq{} and \lq\lq{}software\rq\rq{}) gave a plenary lecture the most memorable feature of which was the introduction of the term \lq\lq{}numerosity\rq\rq{} for the size of a set of data, using letters A, B, C and so on for different sizes.     BS remembers a spirited discussion with Julian where Julian thoroughly debunked this, instead proposing the use of numbers, 1, 2, 3 ... just to say how big the sample was.   

Nevertheless, Julian was very much an advocate for EDA ways of thinking, and when he came back from Princeton he spread the word among the UK statistical community.  Allan Seheult recalls:
\begin{quotation}Julian arrived at Durham from a research post at Princeton, enthusiastic 
about John Tukey's Exploratory Data Analysis and its implementation in 
APL. This was quite distinct from his seminal work in spatial 
statistics, so much so that when he gave talks on EDA at a joint applied 
probability and geography meeting in Bristol and later at the Royal 
Statistical Society conference in Oxford he worried about his reputation 
as an applied probabilist. However, interest in EDA and computing 
illustrated the importance of the analysis of real data to his whole 
approach to statistics. EDA became a distinctive feature of statistics 
courses at Durham, possibly the first such implementation in the UK. 
\end{quotation}
PD recalls a talk he gave to a Royal Statistical Society local group audience jointly with Allan Seheult and Julian.  The topic was median polish, one of Tukey\rq{}s EDA methods which involved smoothing by repeated application of moving medians.  One member of the audience compared the seminar to an episode of \lq\lq{}The Goodies\rq\rq{}, a cult three-hander comedy TV show of the time.  

Julian did not publish much directly on EDA, but his 1981 paper (7) on resistant techniques 
contains many perceptive insights and deserves to be better known.   The first paragraph alone explains very well the issues: 
\begin{quotation}
By a resistant technique will be meant one whose results are at most mildly affected by observations which do not conform to the general pattern of the data. At the outset, it is useful to make a distinction between the application of resistant techniques to exploratory work and to robustness. Consider, for example, a typical regression  problem. A robust analysis usually aims to produce a single set of parameter estimates and associated confidence intervals, just as in classical statistical analysis but employing a procedure which is highly efficient across a wide range of fairly plausible distributional assumptions. However, in exploratory analysis, resistant techniques are used in a more informal manner; often this involves estimation at more than one level of resistance and, when appropriate, a comparison with the results of a standard analysis. In classical and in robust analyses, residuals are important; in exploratory work, residuals are generally of paramount  importance.
\end{quotation}
Julian\rq{}s insistence on integrating EDA techniques within the canon of statistical methods comes out more explicitly later in the paper:  \lq\lq{}... it is necessary to dispel a widespread belief that Tukey\rq{}s approach is intended to replace, rather than augment, careful scientific enquiry.\rq\rq{}   Whether or not that was indeed Tukey\rq{}s view is not for this memoir, but it most certainly was Julian\rq{}s.

The paper deals with resistant approaches in regression analysis, and then goes on to two-way tables, where the links with Julian\rq{}s work on field trials and other classical areas of statistics are clear. Throughout, Julian places EDA in a much wider context.  Also characteristic is his eclectic choice of example data sets, including motorcycle ownership in Great Britain and the repellent effects of lime sulphur on the honeybee.  Finally, in contrast to Tukey\rq{}s multi-coloured pens, Julian effortlessly uses computational approaches, while nevertheless stating \lq\lq{}Median polish has an advantage of simplicity and can often conveniently be carried out by hand.\rq\rq{}  

\section*{Conclusion}

Julian Besag had a reputation for being “difficult,”  but perhaps a better and fairer
adjective would be ``challenging''.  He challenged conventional wisdom. He challenged himself in everything he did, and his accomplishments were remarkable. Most of all, he challenged humbug, whenever and wherever he found it.  When invited to the Statistical Dining Club (associated with the Royal Statistical Society) to celebrate his award of the Society's Guy Medal in Silver, he declined the invitation on the grounds that his wife Valerie was not invited, but never told her about it.  He was, above all, a passionate man -- passionate about his work, passionate about hockey (while he was at Liverpool he had a trial for the Welsh national team) and, in later years when his deteriorating health meant that hockey was no longer an option, passionate about sport in general (Figure \ref{fig:sport}) and, after arrival in Seattle, sailing.

During his years at the University of Washington in the American Pacific North West, Julian
owned five boats, including the one he lived in and the one in his bath but also including  two ocean-going sailboats (see Figures \ref{fig:annie} and  \ref{fig:helm}). After work he would sail up through Puget Sound and beyond. He took beautiful photographs of the islands along the coast between Seattle and Vancouver, and he continued to sail his very large boat single-handedly after suffering kidney failure and having to undertake a punishing regime of self-dialysis, three times a day, every day. 

Julian could be a razor-sharp critic with an unerring eye for a sloppy argument, and when he saw one he was not reluctant to point it out to the perpetrator. But if his criticism of others was sharp, his self-criticism was sharper. Insecure about his lack of formal mathematical training, he repudiated any suggestion that he had mathematical ability, although in truth his understanding was very deep, and he had great perception and originality. He cared deeply about statistics and about his students and his notes were meticulous; on one occasion a lecture didn't go well and so he produced a carefully
written, detailed, printed version which he made available to the
students the very next morning. It is a mark of Julian's generosity and openness to others that he had so many collaborators.  (These have not all been named in this text but they are all in the bibliographies.)   

Noel Cressie, who coincided with Julian at Princeton once wrote:
\begin{quotation}
Julian's research on Markov random fields and spatial point processes was path-breaking. He was my teacher; at times we disagreed about things that seemed small to me. They weren't small to him, and our relationship suffered. I had always hoped for it to improve.
\end{quotation}
Tilmann Gneiting, a colleague from UW days, summed him up thus: 
\begin{quotation}
I'm hopeful Julian would be pleased to know that we recall him as an 
eminent scientist and truth-seeker, a unique character, and in many ways 
the most generous British gentleman. No doubt, Julian's statistical 
legacy as well as our personal memories will continue to inspire 
generations of scientists.
\end{quotation}
His personal relationships -- emotional, social and professional -- were often stormy and very demanding for all concerned, but he was an unforgettable individual, who is much missed\footnote{Memories and tributes from nearly 50 friends and colleagues were received at the time of the memorial meeting and symposium held in April 2011 in Julian's name; they can be seen at 
\texttt{https://www.sustain.bris.ac.uk/JulianBesag/tributes/} or by searching for ``\texttt{Julian Besag tributes}''.}. It was absolutely in character for him, but a great regret to his statistical colleagues, that he refused to allow a celebratory meeting for his 65th birthday, sadly so near the end of his life, because he felt unworthy of it.  
\subsection*{Career summary}

\begin{tabular}{ll}

1968 -- 1969 & Research Assistant to Professor M. S. Bartlett FRS, \\
& Department of Biomathematics, University of Oxford. \\

1970 -- 1975 & Lecturer in Statistics, Department of Computational and Statistical Sciences, \\
& University of Liverpool.\\

1975 -- 1985 & Reader in Mathematical Statistics, Department of Mathematical Sciences,\\
 & University of Durham.\\

1985 -- 1989 & Professor of Statistics, Department of Mathematical Sciences, \\
& University of Durham. \\

1990 -- 1991 & Professor of Statistics, Department of Mathematics and Statistics, \\
& University of Newcastle upon Tyne.\\

1990 -- 2007 & Professor of Statistics, Department of Statistics, \\
& University of Washington, Seattle, U.S.A. \\

2008 -- 2009 & Professor of Statistics, Department of Mathematical Sciences, \\
& University of Bath\\

\end{tabular}

\textbf{Honorary positions}

\begin{tabular}{ll}
1988 -- 1991 & Professor, School of Medicine, University of Newcastle upon Tyne.\\

2002 --  & Consultant, CIMMYT (International Center for Improvement of Maize and Wheat),\\
& Division of Biometry, HQ Mexico City.\\

2006 -- 2008 & Visiting Professor, University of Bath.\\

2007 --  & Professor Emeritus, Department of Statistics, University of Washington.\\

2008 -- 2009  & Research Fellow, School of Mathematics, University of Bristol.\\
\end{tabular}

\subsection*{Service to the scientific community}
\begin{tabular}{ll}
1985--1989 & Member of Council, Royal Statistical Society (and committees).\\
1988--1991 & Chair, NRC/NAS Panel on Spatial Statistics \& Image Processing, U.S.A.\\
1992 & Panel Member, NSF Young Investigators Awards, U.S.A.\\
1992--1996 & Associate Editor, Biometrika.
\end{tabular}

\subsection*{Significant honours and awards}
\begin{tabular}{ll}
1983 & Guy Medal in Silver by the Royal Statistical Society.\\
1984 & Elected Member, International Statistical Institute.\\
1991 & Elected Fellow, Institute of Mathematical Statistics.\\
1992 & Special invited lecture, Institute of Mathematical Statistics.\\
2001 & Chancellor's Medal, University of California, for services to geography.\\
2004 & Fellow of the Royal Society.\\
\end{tabular}

\subsection*{Acknowledgements}
We are very grateful to Julian Besag's aunt Hilde \"{U}berlacker for information about his family background and early life. We thank Valerie Besag, Mandy Chetwynd, Denis Mollison, Debashis Mondal, Bruce Porteous, Debbie Slater and Larissa Stanberry for photographs, and Sheila Bird, Mathias Drton, Liz Green, Julia Mortera, Allan Seheult and Jon Wakefield for helpful conversations and comments.

\newpage
\subsection*{Author profiles}

\begin{center}\emph{Peter Diggle}\end{center}
\parbox[b]{0.3\textwidth}{\includegraphics[width=0.28\textwidth]{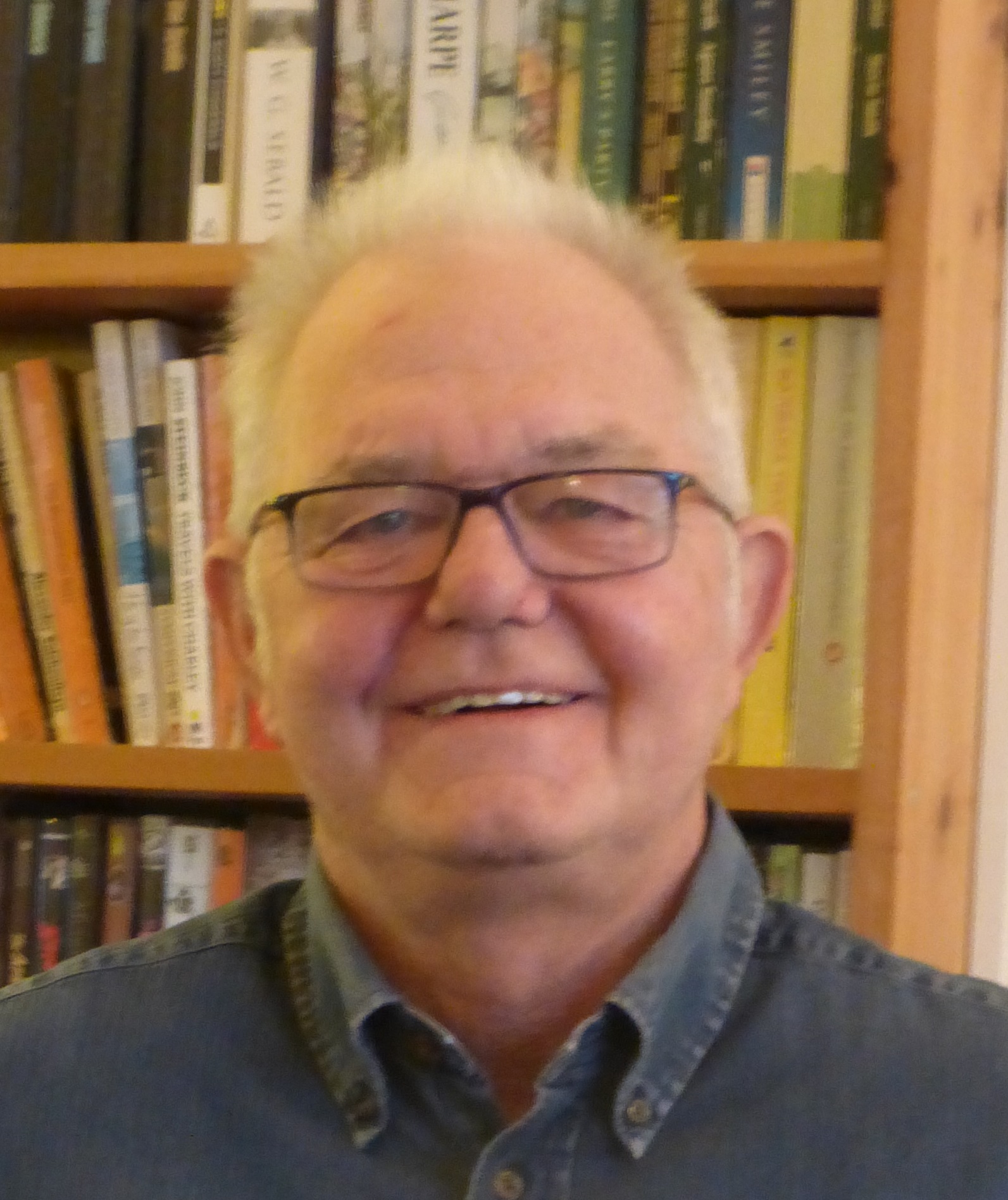}}
\parbox[b]{0.69\textwidth}{Peter Diggle is Distinguished University Professor of Statistics in the Faculty of Health and Medicine, Lancaster University. He also holds adjunct positions at Johns Hopkins, Yale and Columbia Universities, and was president of the Royal Statistical Society from July 2014 to December 2016.  Peter was an undergraduate at the University of Liverpool between 1970 and 1972, where Julian's teaching of statistics and stochastic processes inspired him to embark on a research career.  After taking an MSc in Biomathematics at Oxford, he spent three months back in Liverpool with the intention of completing a PhD under Julian's supervision before taking up a lectureship at the University of Newcastle upon Tyne, with Julian's strong encouragement.}

\begin{center}\emph{Peter Green}\end{center}

\parbox[b]{0.3\textwidth}{\includegraphics[width=0.28\textwidth]{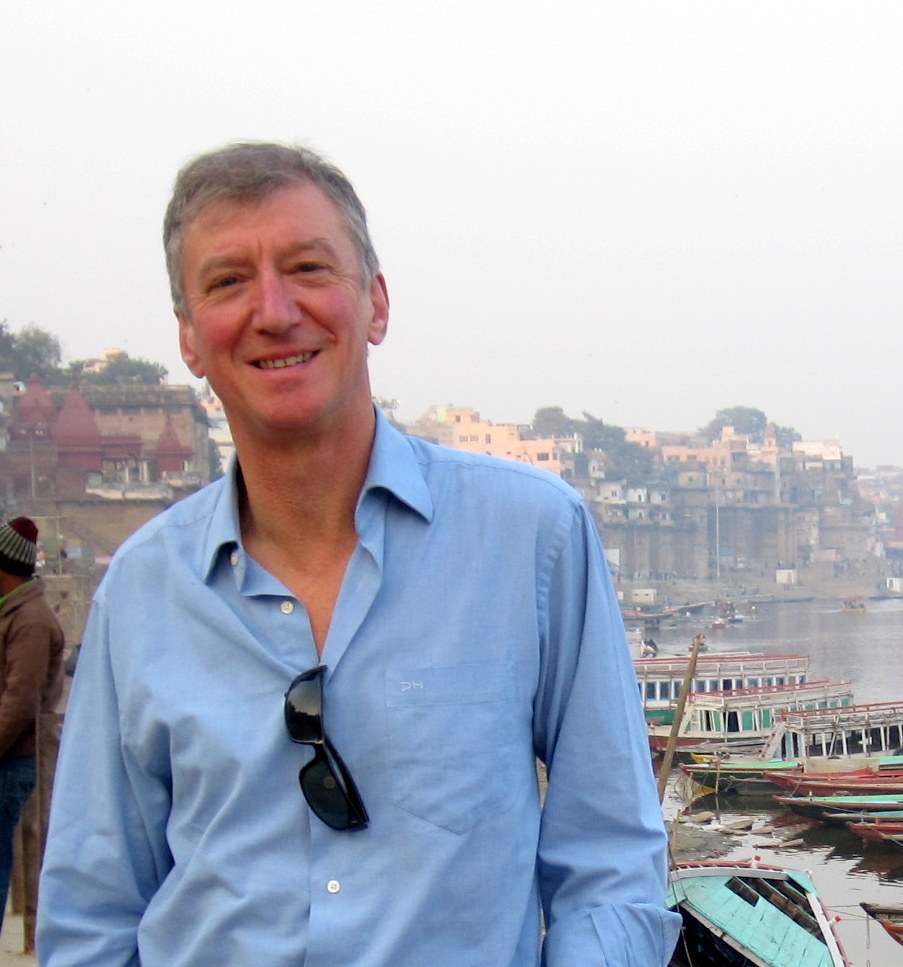}}
\parbox[b]{0.69\textwidth}{Peter Green is Professorial Research Fellow and Emeritus Professor of Statistics at the University of Bristol, and Distinguished Professor of Statistics at the University of Technology, Sydney. He was President of the Royal Statistical Society for 2001--03, and of the International Society for Bayesian Analysis in 2007. He was a junior colleague of Julian Besag at the University of Durham from 1978 to 1989, a period crucial in development of his research career, characterised by intense discussion and collaboration, and occasional joint publication.
This influenced everything he did from then on, giving him a fortunate and happy career, one that Julian supported and encouraged in many ways.
}

\begin{center}\emph{Bernard Silverman}\end{center}

\parbox[b]{0.3\textwidth}{\includegraphics[width=0.28\textwidth]{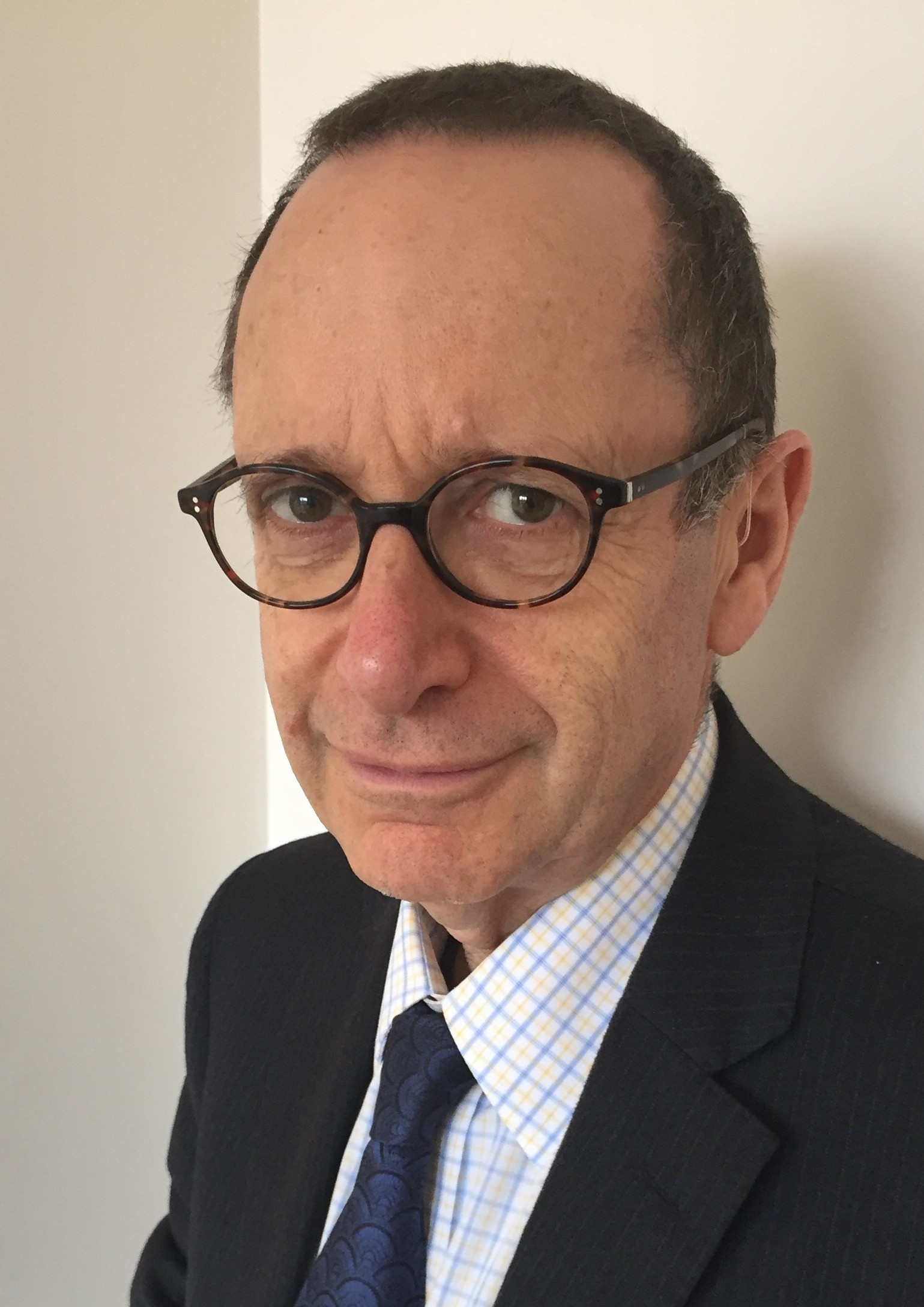}}
\parbox[b]{0.69\textwidth}{Sir Bernard Silverman is Emeritus Professor of Statistics at the Universities of Bristol and Oxford, and former Chief Scientific Adviser to the Home Office.  He is a former president of the Royal Statistical Society and of the Institute of Mathematical Statistics.  He never actually worked in the same institution as Julian at the same time, but was a fellow member of the statistical research community from the mid 1970s to the end of Julian's life.}

\renewcommand\refname{{\large References to other authors}}


\subsection*{Short bibliography}

The following publications are those referred to directly in the text. A full bibliography is available as electronic supplementary material at http://dx.doi.org/xxxx or at \url{http://rsbm.royalsocietypublishing.org}.

\RSreference{1972}{Nearest-neighbour systems and the auto-logistic model for binary data, {\it Journal of the Royal Statistical Society} B {\bf 34}, 75--83.}

\RSreference{1973}{(With J. T. Gleaves). On the detection of spatial pattern
in plant communities. {\em Bulletin of the International Statistical
Institute}, {\bf 45 (1)}, 153--158.}

\RSreference{1974}{Spatial interaction and the statistical analysis
   of lattice systems (with Discussion).
   {\em Journal of the Royal Statistical Society}
   B {\bf 36}, 192--225.}

\RSreference{1975}{Statistical analysis of non-lattice data.
{\em The Statistician}, {\bf 24}, 179--195.\\doi:10.2307/2987782}

\RSreference{1977}{(With P. J. Diggle).  
Simple Monte Carlo tests for spatial pattern. \emph{ Appl.Statist.},
\textbf{26}, 327--333. doi:10.2307/2346974}

\RSreference{1981}{On a system of two-dimensional recurrence relations.
 \emph{Journal of the Royal Statistical Society} B, \textbf{43}, 302--309.}

\RSreference{1981}{On resistant techniques and statistical analysis.  \emph{Biometrika}, \textbf{68},
 463--469.\\doi:10.1093/biomet/68.2.463}

\RSreference{1982}{(with R. Milne and S. Zachary). Point process limits
of lattice processes. {\it Journal of Applied Probability},
{\bf 19}, 210--216. doi:10.1017/S0021900200028424}

\RSreference{1986}{On the statistical analysis of dirty pictures (with
Discussion). {\it Journal of the Royal Statistical Society} B {\bf 48}, 259--302.}

\RSreference{1986}{(With R. A. Kempton). Statistical analysis of field
   experiments using neighboring plots. {\em Biometrics}, {\bf 42}, 231--251. doi:10.2307/2531047}

\RSreference{1989}{Towards Bayesian image analysis. 
 \emph{Journal of Applied Statistics}, \textbf{16}, 395--407. doi:10.1080/02664768900000049}

\RSreference{1989}{(With P. Clifford). Generalized Monte Carlo significance tests.
{\it Biometrika}, {\bf 76}, 633--642. doi:10.1093/biomet/76.4.633}

\RSreference{1991}{(With P. Clifford). Sequential Monte Carlo $p$-values. 
{\it Biometrika}, {\bf 78}, 301--304. doi:10.1093/biomet/78.2.301}

\RSreference{1991}{(With J. Newell). The detection of clusters in rare
   diseases. {\em Journal of the Royal Statistical Society}, A {\bf 154}, 
   143--155.}
	
\RSreference{1991}{(With J. York and A. Molli\'{e}). Bayesian image restoration,
with two applications in spatial statistics (with Discussion).
{\em Annals of the Institute of Statistical Mathematics}, {\bf 43}, 1--59, doi:10.1007/BF00116466}

\RSreference{1993}{(With P. J. Green). Spatial statistics and Bayesian
 computation (with Discussion). {\em Journal of the Royal Statistical Society},
 B, {\bf 55}, 25--37.}

\RSreference{1994}{Discussion of `Representations of knowledge in complex systems' by U. Grenander and M.I. Miller. 
 \emph{Journal of the Royal Statistical Society} B, \textbf{56}, 591--592.}

\RSreference{1995}{(With P. J. Green, D. Higdon and K. L. Mengersen).
 Bayesian computation and stochastic systems (with Discussion).
 \emph{Statistical Science}, \textbf{10}, 3--66.\\doi:10.1214/ss/1177010123}

\RSreference{1995}{(With C. L. Kooperberg).
 On conditional and intrinsic autoregressions.
 \emph{Biometrika}, \textbf{82}, 733--746.}

\RSreference{1998}{(With H. Tjelmeland). Markov random fields with higher-order interactions.  
 \emph{Scandinavian Journal of Statistics}, \textbf{25}, 415--433. doi:10.1111/1467-9469.00113}

\RSreference{1999}{(With D. Higdon). Bayesian analysis of agricultural
field experiments (with Discussion). {\it Journal of the Royal
Statistical Society}, B {\bf 61}, 691--746. doi:10.1111/1467-9868.00201}

\RSreference{2002}{(With F. Bartolucci).
 A recursive algorithm for Markov random fields.
 \emph{Biometrika}, \textbf{89}, 724--730. doi:10.1093/biomet/89.3.724}

\RSreference{2002}{Discussion of `What is a statistical model?' by P. McCullagh.
\emph{The Annals of Statistics}, \textbf{30}, 1267--1277.}

\RSreference{2005}{(With D. Mondal). First-order intrinsic autoregressions
and the de Wijs process. {\it Biometrika}, {\bf 92}, 909--920. doi:10.1093/biomet/92.4.909}

\RSreference{2013}{(With D. Mondal). Exact goodness-of-fit tests for Markov chains. {\it Biometrics}, 69(2), 488-496. doi:10.1111/biom.12009}

\newpage

\end{document}